\documentclass[aps,prl,twocolumn,groupedaddress,showpacs]{revtex4}
\usepackage{graphicx}
\usepackage{amsmath}

\begin{document}

\title{Observation of shock waves in a strongly interacting Fermi gas}

\author{James Joseph$^{1}$}
\author{John Thomas$^{1}$}
\affiliation{$^{1}$Department of Physics, Duke University, Durham, North Carolina, 27708, USA}
\author{Manas Kulkarni$^{2,3}$}
\author{Alexander G. Abanov$^{2}$}
\affiliation{$^{2}$Department of Physics and Astronomy, Stony Brook University,
Stony Brook, NY 11794-3800, USA}

\affiliation{$^{3}$Department of Condensed Matter Physics and Material Science,
Brookhaven National Laboratory, Upton, NY-11973, USA}


\date{\today}

\begin{abstract}
We study collisions between two strongly interacting atomic Fermi gas clouds. We observe exotic nonlinear hydrodynamic behavior, distinguished by the formation of a very sharp and stable density peak as the clouds collide and subsequent evolution into a box-like shape.  We model the nonlinear dynamics of these collisions using quasi-1D  hydrodynamic equations. Our simulations of the time-dependent density profiles agree very well with the data and provide clear evidence of shock wave formation in this universal quantum hydrodynamic system.
\end{abstract}

\maketitle

Shock waves are of broad interest and can occur in nonlinear quantum hydrodynamic systems when regions of high density move with a faster local velocity than regions of low density resulting in large density gradients. In the absence of dissipative or dispersive forces large gradients would eventually lead to a ``gradient catastrophe.''  In cold atomic gases, quantum shock waves have been observed by using slow light methods to create sharp density features in a weakly interacting Bose-Einstein condensate (BEC)~\cite{Dutton}. Predictions of the density profiles for dispersive shock waves propagating in weakly interacting BEC's agree with observations, while predictions for dissipative classical shock waves only reproduce the slowly varying features of the density profiles~\cite{Hoefer}.  In BEC's, dispersive shock waves produce soliton trains~\cite{Dutton}, which also have been observed and modeled for rapidly rotating BEC's~\cite{Simula} and for merging and splitting BEC's~\cite{Chang}. While quantum shock waves have been studied theoretically in BEC's~\cite{Hoefer,Damski,Kamchatnov1}, they are also of recent interest in theories of non-equilibrium electron Fermi gases~\cite{Abanov,Abanov1}.

In this Letter, we show that strongly interacting atomic Fermi gases provide a new universal medium for studies of nonlinear hydrodynamics of quantum matter. Strong interactions are characterized by a divergent s-wave scattering length, which is obtained by using a bias magnetic field to tune to a broad collisional (Feshbach) resonance. In this so-called unitary regime, the chemical potential and the pressure are universal functions of the density and temperature.  In common with a quark-gluon plasma~\cite{PhysicsToday}, a state of matter that existed microseconds after the Big Bang, strongly interacting Fermi gases exhibit features of nearly perfect (low viscosity) hydrodynamics, such as anisotropic (elliptic) flow~\cite{OHaraScience}, and serve as a paradigm for other exotic quantum hydrodynamic Fermi systems, such as nuclear matter~\cite{Bjorken}.

Our experiments employ a 50:50 mixture of the two lowest hyperfine states of $^6$Li, confined in a cigar-shaped CO$_2$ laser trap, and bisected by a blue-detuned beam at 532 nm, which produces a repulsive potential. The gas is then cooled via forced evaporation near a broad Feshbach resonance at 834 G~\cite{BartensteinFeshbach}.  After evaporation, the trap is adiabatically recompressed to 0.5\% of the initial trap depth. This procedure produces two spatially separated atomic clouds, containing  a total of  $\simeq 10^5$ atoms per spin state.    In the absence of the blue-detuned beam, the trapping potential is cylindrically symmetric with a radial trap frequency of $\omega_x=\omega_y=\omega_\perp=2\pi\times 437 $~Hz and an axial trap frequency of $\omega_z = \sqrt{\omega_{Oz}^2+\omega_{Mz}^2}=2\pi\times\, 27.7$~Hz, where the axial frequency of the optical trap  is $\omega_{Oz}=2\pi\times 18.7$~Hz and $\omega_{Mz}=2\pi\times\, 20.4$~Hz arises from curvature in the bias magnetic field.  When the repulsive potential is abruptly turned off, the two clouds accelerate toward each other and collide in the CO$_2$ laser trap. After a chosen hold time, the CO$_2$ laser is turned off, allowing the  atomic cloud to expand for 1.5 ms, after which it is destructively imaged with a  5 $\mu$s pulse of resonant light.

\begin{figure*}[htb]
\includegraphics[width=5.0 in]{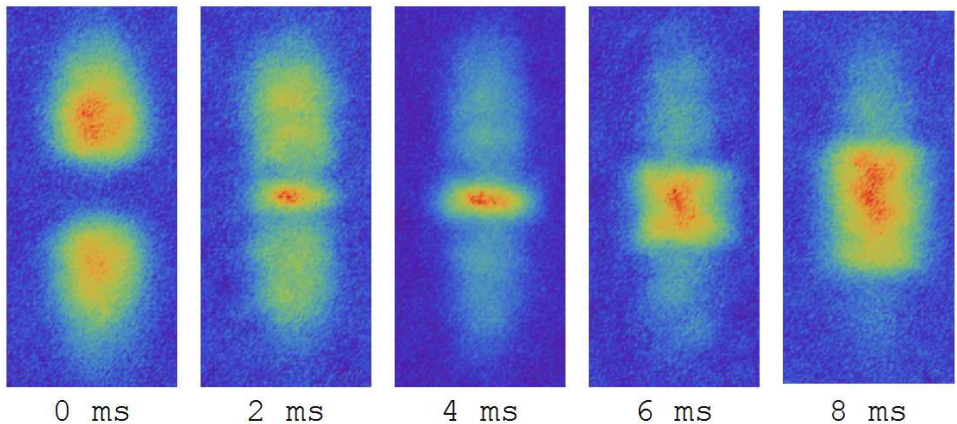}
\caption{\label{fig:Images}  Collision between two unitary Fermi gas clouds in a cigar-shaped optical trap. The clouds are initially separated by a repulsive 532 nm optical beam. After the 532 nm beam is extinguished (0 ms), the clouds approach each other.  False color absorption images show the spatial profiles versus time. Initially, a sharp rise in density occurs in the center of the collision zone.  At later times the region of high density evolves from a ``peak-like" shape into a ``box-like" shape. The well defined edges of the central zone in the last three images provide evidence of shock wave formation in the unitary Fermi-gas.}
\end{figure*}

Fig.~\ref{fig:Images} shows false color absorption images for a collision of the atomic clouds at different times after the blue-detuned  beam is extinguished.  Two distinctive features are clearly seen in this data: (i) the formation of a central peak, which is well-pronounced and robust; (ii) the evolution of this peak into a box-like shape with very sharp boundaries.  The observed large density gradients provide strong evidence of shock wave formation in this system, where the sharp boundaries of the ``box" are identified as shock wave fronts.  Numerical modeling of the hydrodynamic theory for one dimensional motion is used to predict the evolution of the atomic density, yielding profiles in good agreement with the data.

For simplicity, we assume that the cloud is a unitary Fermi gas at zero temperature, i.e., we model the cloud as a single fluid,  consistent with our measurements of the sound velocity~\cite{JosephSound}. In this case, the local chemical potential has the universal form  $\mu(n_{3D}) = (1+\beta)\epsilon_F(n_{3D})$, where $\epsilon_F(n_{3D}) = \frac{\hbar^2}{2m}(3\pi^2n_{3D})^{2/3}$ is the ideal gas local Fermi energy corresponding to the three-dimensional density $n_{3D}$.  Here,  $\beta=-0.61$ is a universal scale factor~\cite{OHaraScience,Heiselberg,ThermoLuo}.

Neglecting viscous forces,  the dynamics for the density $n_{3D}(\mathbf{r},t)$ and the velocity field $\mathbf{v}(\mathbf{r},t)$,  are described by the continuity equation,
\begin{equation}
\label{eq:cont3D}
\partial_{t}n_{3D}+\mathbf{\nabla}\cdot\left(n_{3D}\mathbf{v}\right)=0
\end{equation}
and the Euler equation,
\begin{equation}
 \label{eq:euler3D}
	m\partial_{t}\mathbf{v}+\mathbf{\nabla}\left[\mu(n_{3D})+U_{trap}(r,z)
	+\frac{1}{2}m\mathbf{v}^2\right]=0,
\end{equation}
where we assume irrotational flow. Here  $U_{trap}(\mathbf{r})=\frac{1}{2}m\omega_\perp^2r^2+\frac{1}{2}m\omega_z^2z^2$ is the confining harmonic potential of the cigar-shaped trap.

To determine the initial density profile for the separated clouds, we consider the equilibrium 3D density of the Fermi gas in the trap, including a knife-shaped repulsive blue-detuned beam potential $V_{rep}(z)$.  The blue-detuned laser beam is shaped by a cylindrical lens telescope, so that the spot size is small compared to the long dimension of the cigar-shaped cloud and large compared to the transverse dimension.  Therefore, the repulsive potential can be considered to vary only in the z (axial) direction according to $V_{rep}(z) = V_0 \exp\left(-(z-z_0)^2/\sigma_z^2\right)$. We measure the width  $\sigma_z=21.2\,\mu$m. The offset $z_0=5\,\mu$m of the focus from the center in the long direction of the optical trap is determined  by a fit to the first density profile at 0 ms.   Using the beam intensity and the ground state static polarizability of $^6$Li at 532 nm, we find $V_0=12.7\,\mu$K.  The initial density profile is then

\begin{equation}
 \label{EQ:3DDensityWV}
	n_{3D}(r,z) =\tilde{n}\left(1-\frac{r^2}{R_{\perp}^2}-\frac{z^2}{R_z^2}
	- \frac{V_{rep}(z)}{\mu_G}\right)^{\frac{3}{2}},
\end{equation}
where $\tilde{n}= [(2m\mu_G/\hbar^2)/(1+\beta)]^{3/2}/(3\pi^2)$.
In Eq.~\ref{EQ:3DDensityWV}, $R_{z,\perp}=\sqrt{2\mu_G/(m\omega_{z,\perp}^{2})}$ and $\mu_G$ is the global chemical potential, which is determined by normalizing the integral of the 3D density to the total number $N$ of atoms in both spin states. For $N=2\times 10^5$, we find $\mu_G = 0.53\,\mu$K, $R_z=220\,\mu$m, and $R_\perp=14\,\mu$m.

We note that $\mu_G/(\hbar \omega_\perp)=27$, which means that the typical number of filled energy levels of transverse quantization is large. Therefore, in this paper we use 3D hydrodynamics, Eqs.~\ref{eq:cont3D}~and~\ref{eq:euler3D}, and neglect effects of transverse quantization even though they are more pronounced in regions with lower density.

We model the dynamics for the one-dimensional motion in the long direction of the cigar-shaped trap. Just after the blue-detuned beam is extinguished, the initial 1D density profile is determined by integrating $n_{3D}$ of Eq.~\ref{EQ:3DDensityWV} over the transverse dimension $r$,
\begin{equation}
 \label{EQ:1DDensityWV}
	n_{1D}(z) = \frac{2 \pi}{5} R_\perp^2 \tilde{n}\left(1-\frac{z^2}{R_z^2}
	- \frac{V_{rep}(z)}{\mu_G}\right)^{\frac{5}{2}}.
\end{equation}

In the following we assume that during the evolution the $r$ dependence of Eq.~\ref{EQ:3DDensityWV} is preserved with the effective size of the cloud being a slow function of $z$ and $t$. We also assume that the hydrodynamic velocity is along $z$ axis and does not depend on $r$. Then the subsequent time evolution of the density follows the quasi-1D nonlinear hydrodynamic equations:
\begin{align}
\label{eq:cont}
\partial_{t}n&=-\partial_{z}\left(nv\right) \\
 \label{eq:euler}
\partial_{t}v&=-\partial_{z}\left(\frac{v^2}{2}+Cn^{\frac{2}{5}}+\frac{1}{2}\omega_z^2z^2\right) +\nu\frac{\partial_{z}(n\partial_{z}v)}{n},
\end{align}
where $C=\frac{1}{2}\omega_\perp^2 l_\perp^2\left(\frac{15\pi}{2} l_\perp\right)^{2/5}(1+\beta)^{3/5}$ and $l_\perp=\sqrt{\hbar/(m\omega_{\perp})}$ is the oscillator length. For brevity, we have omitted the subscript $1D$ in Eqs.~\ref{eq:cont}~and~\ref{eq:euler}.  The last ``viscosity" term in Eq.~\ref{eq:euler} is added \textit{phenomenologically} to describe dissipative effects. For the unitary 1D  fluid,  $\nu$ is the effective kinematic viscosity, which has a natural scale $\hbar/m$. It is the only  fitting parameter in the theory~\cite{Viscosity}.

For sound wave experiments with a small pulse $V_0$, one can linearize the differential equations (\ref{eq:cont}) and (\ref{eq:euler}) around an equilibrium density configuration $n_{0}(z)$ in a harmonic trap. Defining $n(z,t)\equiv n_0(z)+\delta n(z,t)$, the linearized evolution equation for $\delta n(z,t)$ (neglecting viscosity) is
\begin{equation}
 \label{cwe}
	\partial_t^2\delta n=\partial_z\left[n_0
	\partial_z\left(\frac{2C}{5m}n_0^{-\frac{3}{5}}\delta n
	\right)\right].
\end{equation}
For a flat background density, i.e., constant $n_0$, with $\mu_G=Cn_0^{2/5}$,   Eq.~\ref{cwe} reduces to $\partial_t^2\delta n=c^2\partial_z^2\delta n$ with the sound velocity $c=\sqrt{2 \mu_G/5 m}$, in agreement with previous theory~\cite{CapuzziSFSound,bertaina} and experiment~\cite{JosephSound}.

To compare the numerical solutions of Eqs.~\ref{eq:cont}~and~\ref{eq:euler} with experiment, we note that the images are taken after an additional free expansion for $1.5\mbox{ ms}$, during which $n_{1D}$ continues to slowly evolve in the axial potential of the bias magnetic field, i.e., $\omega_z\to\omega_{Mz}=2\pi\times 20.4$~Hz. We assume that during this expansion, the transverse density profiles keep the same form, but the radius increases with time. Then $n_{3D}(r,z)\to n_{3D}(r/b_\perp,z)/b_\perp^2$, where $b_\perp(t)$  is a transverse scale factor, which obeys $\ddot{b}_\perp=\omega_\perp^2\,b_\perp^{-7/3}$, with $b_\perp(0)=1$ and $\dot{b}_\perp(0)=0$~\cite{OHaraScience,StringariExpansion,StringariReview}. Since the 3D pressure scales as $n_{3D}^{5/3}$, the 1D pressure scales as $b_\perp^{-4/3}$. This leads to a simple modification of Eq.~\ref{eq:euler}: $C\to C(t)=C/b_\perp^{4/3}(t)$.

\begin{figure}[h]
\includegraphics[width=3.00 in]{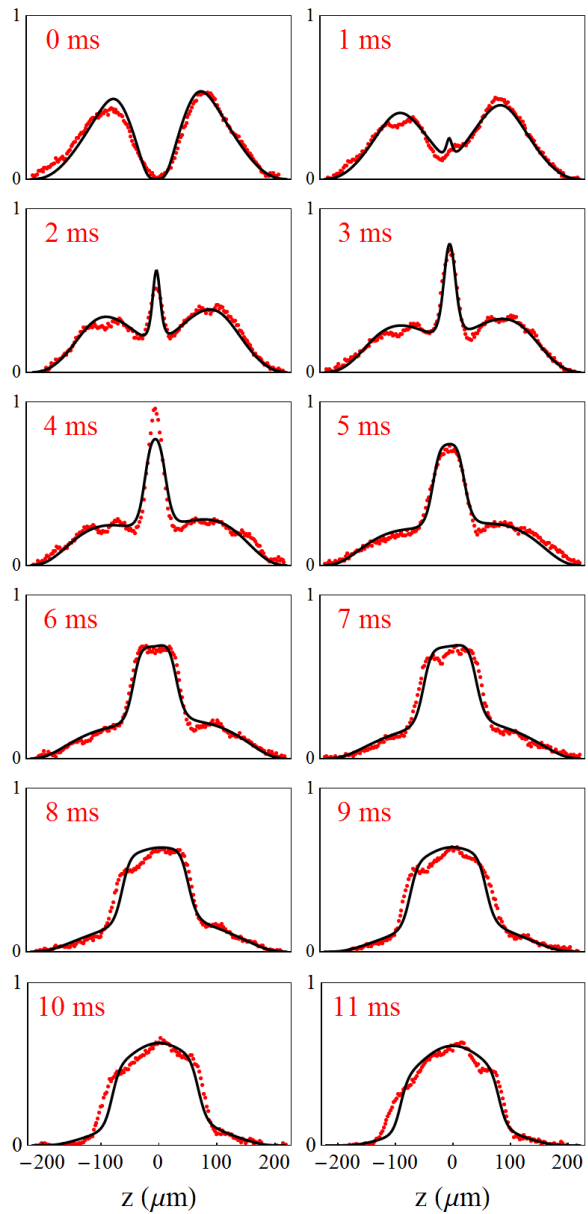}
\caption{\label{fig:comparison} 1D density profiles divided by the total number of atoms versus time for two colliding unitary Fermi gas clouds.  The normalized density is in units of  $10^{-2}/\mu$m per particle. Red dots show the measured 1D density profiles. Black curves show the simulation, which uses the measured trap parameters and the number of atoms, with the kinetic viscosity as the only fitting parameter.}
\end{figure}

We numerically integrate Eqs.~\ref{eq:cont}~and~\ref{eq:euler} using the measured values of the trap frequencies, atom number, and the offset, depth, and width of the repulsive potential.  In the numerical simulation we create and load a density array as well as a velocity array with grid spacing $\delta_z$.  The initial velocity is set to zero. The simulation then updates the density and velocity field in discrete time steps $\delta_t$ according to Eqs.~\ref{eq:cont}~and~\ref{eq:euler}.  The 1D density profiles are calculated as a function of time after the repulsive potential is extinguished.  Fig.~\ref{fig:comparison} shows the predictions and the data, which are in very good agreement.  For the simulation curves shown in the figure we use a grid of 150 points. To check for numerical consistency, we also employ a smoothed-particle-hydrodynamics~\cite{sph} approach, where the fluid is described by discrete pseudo-particles.  The results obtained indeed coincide with the discretized-grid approach described above.

As shown in Fig.~\ref{fig:comparison}, we observe a dramatic evolution for the density of the gas.  During the collision, a distinct and stable density peak forms at the point of collision in the center of the trap~\cite{DipSplitting}.  The density gradient at the side of the central peak increases from its onset until $\approx 3$~ms, at which point the gradient reaches its maximum value.  A large gradient at the edge of the collision zone is maintained throughout the rest of the experiment. For most of the data, we find relatively small deviations from the simulation. The largest deviation occurs at 4 ms, where the maximum density of the observed central peak exceeds that of the simulation by $\simeq 20$\%.

The steep density gradients observed in Fig.~\ref{fig:Images} suggest shock wave formation.  A deeper analysis of the simulation curves provides additional evidence for shock waves. Without any dissipation, the numerical integration of the quasi-1D theory breaks down due to a ``gradient catastrophe." We find that the dissipative force in Eq.~\ref{eq:euler}, which is described by the kinematic viscosity coefficient $\nu$, is required to attenuate the large density gradients and avoid gradient catastrophe.  For the data shown in Fig.~\ref{fig:comparison}, we find that the best fits are obtained with the viscosity parameter $\nu=10\,\hbar/m$.  For smaller values of $\nu$, the simulation produces qualitatively similar results to those shown in the figure, only with steeper density gradients at the edges of the collision zone.  The dissipative term  $\propto\nu$ has a relatively small effect on the density profiles, unless we are in a shock wave regime, where the density gradients are large.  Hence, the numerical model suggests that the large density gradient observed at the edge of the collision zone is the leading edge of a dissipative shock wave.

Our data for a strongly interacting Fermi gas are very well described by a simple one-dimensional model based on dissipative nonlinear quantum hydrodynamics. In this sense, a unitary Fermi gas is different than a weakly interacting Bose-Einstein condensate, where dispersive terms play a major role.

We derive the one dimensional model, with an effective one-dimensional chemical potential $\mu_{1D}=C\,n_{1D}^{2/5}$, assuming a single fluid near the ground state. However, we expect that at higher temperatures,  even in the normal fluid regime, rapid collisional equilibrium in the unitary gas will produce nearly adiabatic evolution with a three-dimensional pressure $\propto n^{5/3}$, and hence an identical power-law dependence for the effective one-dimensional chemical potential.

Our observations suggest that collisions of two clouds of ultra-cold atoms in the unitary regime are accompanied by shock wave formation. In future work, it will be interesting to study the origin of the effective viscosity and the effects of transverse quantization. Further, the large density gradients produced in the experiments suggest that it may be possible to investigate the effects of higher derivative dispersive terms in the stress tensor~\cite{Abanov,meppilink,Abanov1,Hoefer}. Finally, the radial density variations observed in the two dimensional image is not captured in the one-dimensional profiles, but may be studied by expanding the analysis to higher dimensions.

The work of the Duke group is supported by the Physics Divisions of the  National Science Foundation, the Army Research Office, the Air Force Office Office of Sponsored Research, and the Division of Materials Science and Engineering,  the
Office of Basic Energy Sciences, Office of Science, U.S. Department of Energy. The work of A. G. A. was supported by the NSF under Grant No. DMR-0906866. We are grateful to P. Wiegmann, E. Shuryak, D. Schneble and F. Franchini for useful discussions.


\begin{thebibliography}{24}
\bibitem[{Vis()}]{Viscosity}
\bibinfo{note}{In a unitary Fermi gas, which is scale invariant, the bulk
  viscosity vanishes. We expect that $\nu$ arises from shear viscosity, which
  has a natural scale $\eta\propto\hbar\,n$, so that $\nu=\eta/(nm)\propto
  \hbar/m$. See Ref.~\cite{CaoViscosity}.}
\bibitem[{Dip()}]{DipSplitting}
\bibinfo{note}{In linear hydrodynamics one also observes a peak formation,
  which can be interpreted as a residual background in the middle after a small
  initial dip splits into left and right moving dips (observed in
  Ref.~\cite{JosephSound}).}
\end{thebibliography}

\end{document}